%% file: main.tex
\documentclass[12pt,aps,prx,english,superscriptaddress,floatfix,longbibliography]{revtex4-2}
\usepackage[utf8]{inputenc}

\usepackage{amssymb}  % for fancy Z2 in references, LOL

\usepackage{lineno} %% Add line numbers w/ \linenumbers.

\newcommand{\reportnumber}{FERMILAB-PUB-22-437-QIS,IQuS@UW-21-028}
\usepackage[angle=0,scale=1,color=black,firstpage=true,opacity=1]{background}

\SetBgContents{\small\texttt{\reportnumber}}
\SetBgPosition{current page.north east}
\SetBgHshift{-6.5cm}
\SetBgVshift{-1.5cm}
\begin{document}
%\linenumbers

%% May 31, 2022: Preliminary reports by the Topical Groups
\title{Snowmass Computational Frontier: Topical Group Report on Quantum Computing} %%CompF6

%% This is copyright licensing statement from DOE
\thanks{The United States Government retains and the publisher, by accepting the article for publication, acknowledges that the United States Government retains a non-exclusive, paid-up, irrevocable, world-wide license to publish or reproduce the published form of this manuscript, or allow others to do so, for United States Government purposes. The Department of Energy will provide public access to these results of federally sponsored research in accordance with the DOE Public Access Plan. (http://energy.gov/downloads/doe-public-access-plan)}
\date{1 July 2022}

\author{Travis S.~Humble}
\affiliation{Quantum Science Center, Oak Ridge National Laboratory, Oak Ridge, Tennessee 37830 USA}

\author{Gabriel N. Perdue}
\affiliation{Fermi National Accelerator Laboratory, Fermilab Quantum Institute, PO Box 500, Batavia, Illinois, 60510-0500 USA}
%\orcid{}

\author{Martin J.~Savage}
%\email{mjs5@uw.edu}
\affiliation{InQubator for Quantum Simulation (IQuS), Department of Physics, University of Washington, Seattle, Washington 98195 USA}

\begin{abstract}

Quantum computing will play a pivotal role in the High Energy Physics (HEP) science program over the early parts of the 21$^{st}$ Century, both as a major expansion of our capabilities across the Computational Frontier, and in synthesis with quantum sensing and quantum networks.  
This report outlines how Quantum Information Science (QIS) and HEP are deeply intertwined endeavors that benefit enormously from a strong engagement together.
Quantum computers do not represent a detour for HEP, rather they are set to become an integral part of our discovery toolkit.
Problems ranging from simulating quantum field theories, to fully leveraging the most sensitive sensor suites for new particle searches, and even data analysis will run into limiting bottlenecks if constrained to our current computing paradigms. 
Easy access to quantum computers is needed to build a deeper understanding of these opportunities.
In turn, HEP brings crucial expertise to the national quantum ecosystem in quantum domain knowledge, superconducting technology, cryogenic and fast microelectronics, and massive-scale project management.
The role of quantum technologies across the entire economy is expected to grow rapidly over the next decade, so it is important to establish the role of HEP in the efforts surrounding QIS.
Fully delivering on the promise of quantum technologies in the HEP science program requires robust support.
It is important to both invest in the co-design opportunities afforded by the broader quantum computing ecosystem and leverage HEP strengths with the goal of designing quantum computers tailored to HEP science. 

\end{abstract}
\maketitle

\tableofcontents
\newpage

%%%%%%%%%%%%%%%%%%%%%%%
\section{Executive Summary}
\label{sec:newexecsummary}
\input{new_executive_summary}

%%%%%%%%%%%%%%%%%%%%%%%
\section{Reccomendations}
\label{sec:recommendations}
\input{recommendations}

%%%%%%%%%%%%%%%%%%%%%%%
\section{Introduction}
\label{sec:execsummary}

The science program of High Energy Physics (HEP) stands to benefit tremendously from current and future Quantum Information Science (QIS) technologies and applications.
QIS is viewed through a lens that defines three major technologies --- quantum computing, quantum sensors, and quantum networks.
In practice, these three areas often blend together.
The distinguishing characteristic of ``quantum'' science and technology is the 
utilization of entanglement and coherence, inherent features of quantum mechanics that are not leveraged by ``classical'' technologies.
Quantum computing uses entanglement and superposition as algorithmic primitives in algorithms.
This enables a programmable interface to quantum physics, enabling experiments and computations that are not possible in any other way.
Because entanglement and superposition are woven into the fabric of reality at the deepest level, it should not be surprising that quantum concepts represent powerful enabling technological advances, and
the continued integration of the unique expertise and infrastructure within HEP with QIS research priorities is essential.
While there are strong overlaps with quantum sensors and networks, here we focus on the application of quantum technology to solving computational problems.

This is the first {\it Snowmass process} to consider QIS, and given its anticipated large impact on our research activities in the coming decade, this is a very important, formative time.
To provide some context for this report, as of today, modern quantum computers are classified as being ``Noisy, Intermediate-Scale Quantum'' (NISQ) \cite{Preskill_2018} machines and distinguished by the characteristics of being  noisy (suffer from decoherence) and featuring O(100) qubits.
By the end of the time frame considered by this Snowmass planning process, many anticipate a shift in this paradigm to an era of scalable quantum error correction (QEC) and fault-tolerant quantum computing, which is widely expected to open new frontiers for computation far beyond what would be achieved with conventional computing alone. These advances are powered, in part, by algorithms with polynomial, or in some cases exponential, scaling advantages over the best known classical approaches.
Therefore, today's research is fundamentally exploratory and developmental in nature but valuable because of the anticipated future acceleration in performance. The direction set by the Snowmass planning process will have a profound impact on what can be accomplished.

Throughout this report, reference will be made to HEP involvement in QIS and vice versa, considering them as separate domains.
To some extent of course they are --- QIS stands as its own discipline and HEP has 
developed technologies for many years without explicit connection to QIS.
While it is taxonomically convenient to treat them separately, the two disciplines are intertwined at a fundamental level:  
QIS is deeply connected to the fundamental physics questions of HEP just as the broader area of HEP research brings significant value to the national quantum ecosystem.
More specifically, HEP brings expertise in large-scale simulation, superconducting devices and materials, fast and cryogenic electronics, domain expertise in quantum physics, science driver applications that motivate the development of quantum computers, experience with large and complex projects, and an important role in the national science and technology workforce development chain.
In many cases HEP science applications are fertile new ground for research by the QIS community and, analogously to the evolution of  high-performance computing, expected to be drivers for developing quantum technologies, algorithms, and software.
It is useful to think of quantum computers as providing a programmable experimental platform for quantum physics.
In this sense, they are discovery tools like telescopes or particle accelerators.

During the next decade and beyond, 
HEP faces an array of computational challenges where the only practical path(s) to solution requires the utilization of entanglement and superposition.
As first identified by Richard Feynman and others in what are considered to be the foundational papers in quantum computing~\cite{5392446,5391327,Manin1980,Benioff1980,Feynman1982,Fredkin1982,Feynman1986,doi:10.1063/1.881299}, %% do we really need these?
scalable methods for accurate simulations of quantum many-body systems are beyond the capabilities of classical computation.
The last decade has witnessed these limitations becoming increasingly obvious within areas of HEP research.

The development and integration of robust quantum simulation capabilities into the broader HEP lattice gauge theory program is essential to address 
processes and systems that cannot be accessed experimentally and/or with sufficient precision.  These include real-time dynamics for more precisely including quantum coherence and entanglement in event generation and fragmentation, and in realizing the matter-anti-matter asymmetry generated in the early universe. Additionally, quantum computers are anticipated to be required for a selection of simulation tasks in support of event generators, from accelerating matrix-element calculations to enabling HEP-relevant nuclear physics calculations, such as  neutrino-nucleus scattering cross section calculations involving large nuclei --- 
relevant calculations of nuclei beyond carbon with the necessary precision are daunting.
For these HEP challenges, and many more outside of HEP,  error-corrected quantum computers could far outpace what can be achieved using classical computers, creating  opportunities for more broadly defined discovery science using quantum computers.

The multi-faceted aspect of quantum simulation, and the limited fidelity and circuit-depth capabilities of present devices dictates that the quantum architectures that are optimal for HEP applications in the future have not yet been identified. 
A variety of different technologies are under consideration including qubit-architectures based on trapped-ions, superconducting qubits, cold-atoms (Rydberg), and optical systems.  
Qutrit and qudit architectures have also been explored, for instance, in the context of trapped ions and superconducting radio-frequency (SRF) cavities.
The significance of purely analog and hybrid (digital-analog) simulations is also being explored, for example, by using cold atom system as an analog frameworks for simulating synthetic gauge fields.
Current analyses of the available digital and analog qubit and qudit computing systems have begun to quantify the quantum resources required for the road ahead, and present estimates of required resources are likely pessimistic.
Co-design efforts during the next decade will be critical to advancing the HEP mission, by identifying the implementing algorithms, protocols and specialized hardware to accelerate HEP-centric components of quantum simulation. 
These developments will require a public-private partnership spanning multiple stakeholders, but the unique HEP capabilities and expertise focused toward addressing HEP challenges are assured to accelerate progress toward multiple QIS-specific objectives.
These efforts have been undertaken in collaborations between national laboratories, technology companies and universities, involving diverse small to modest size teams.
They are analogous to the Scientific Discovery through Advanced Computing (SciDAC \footnote{\protect\url{https://www.scidac.gov}}) teams that 
utilize classical computing for scientific applications, and involve a range of complex and diverse activities, from the science-level application code down to the device-level instructions.

An exciting major effort emerging from within HEP is one of the US Department of Energy National Quantum Information Science Research Centers (NQISRCs) hosted at Fermilab. The Superconducting Quantum Materials and Systems (SQMS) Center is utilizing the SRF-cavities that play a critical role in particle accelerators for designing and constructing multiply-connected, large local-Hilbert space quantum devices. Arising from a coordinated partnership of government, industry and academia, these designs are anticipated to be well-suited for addressing lattice problems, including HEP-specific lattice quantum field theories.  
These focused co-design efforts must exist alongside synergistic research in quantum simulation design and execution employing cloud-accessible or locally accessible quantum devices.
%Current and new partnerships with technology companies and start-ups, universities and national laboratories are important for implementing these strategies.
As the field of quantum computing matures moves from the NISQ era to the era of fault-tolerant computing, we will then be prepared with a comprehensive strategy to asses the viable and optimal paths to address HEP simulation challenges. 
Of course, SQMS is not the only NQISRC and it is not the only Center where HEP scientists are playing a role. 
For example, the Quantum Science Center (QSC) is another place where HEP science is a priority and where HEP expertise is advancing problems across a broad scientific spectrum.

Moving beyond simulation, quantum computers also have the potential to impact data analysis in HEP.
There is great interest in the potential of quantum computers as computational accelerators for machine learning (ML) and optimization tasks.
While it is likely these applications will require quantum error-correction, it is important to invest in the research needed for the HEP community is able to leverage these unique methods when they become available.
HEP analyses are expected to become increasingly impacted by  quantum data --- for example, from quantum sensors and the output of quantum simulations --- we are in a position to drive research on quantum machine learning (QML) and data analysis tasks from this early stage. Our applications typically benefit earlier than other commercial entities, and
this provides unique and interesting opportunities for new partnerships that could benefit the broader HEP science program.

Large-scale classical computing resources, spanning from cloud-accessible capacity to leadership-class capability, are required to advance the quantum frontier within HEP in sensing, communication and simulation.
Although the exact simulation of quantum computers and quantum devices scales exponentially and outstrips the entire global computing cloud once we reach a scale of $O(50)$ qubits, there are a large number of practical applications that would like to run at the largest resource scale possible.
For example, there are important needs for materials design for quantum sensing experiments, and for open system simulation packages to support efficient quantum gate design, among many other applications.
Over the course of the next decade, we expect to move into the era of quantum error correction --- this will enable computers capable of performing calculations that are expected to robustly have  capabilities beyond classical computers.
However, the development of error-corrected quantum computers will require massive classical simulations, expected to stress Leadership Class Computational Facilities, and produce extremely large datasets.
For example, full density-matrix simulations of N-qubit systems
requires tracking data objects that scale as $2^N \times 2^N$, and simulating quantum noise means it is not possible to treat these objects as sparse matrices.
Truncations that limit simulated entanglement will be required, and there will essentially always be a demand for more faithful simulations which require more memory and more storage.
Therefore, quantum research is expected to make demands on storage and compute in aggregate that are comparable to a particle physics experiment, and proper investments in tooling and workflow support should not be overlooked.

Beyond the classical resources required for direct simulation of devices~\cite{opensystemsim}, there are massive classical resources required for theoretical and algorithm developments directly related to HEP simulation, that in part have been enabled by progress in QIS. 
Tensor network methods are now playing an increasingly important role in HEP and QIS as they enable reformulations of lattice models 
that are suitable for quantum simulation and coarse-graining methods. Importantly, tensor networks can also be used to perform classical simulations of quantum circuits, providing an important capability for advancing QIS through developing and testing quantum computing algorithms.
Tensor methods also provide tools to quantify entanglement in conformal field theories, with their connections to gravity.  
Tensor networks also have interesting applications in machine learning, especially quantum machine learning.
There is a growing interdisciplinary community in this area, in which HEP plays a prominent role. 

The co-design of classical processing and control hardware is likely to play a key role in advancing elements of the HEP quantum program, enabled by, for example, FPGA and ASIC micro-electronics arrays.
We envisage a growing demands for such capabilities within HEP, beyond that required for artificial intelligence and machine learning.
This is also a key area where HEP expertise is valuable to the national ecosystem, and investments are highly likely to offer very broad, positive economic impact.
For example, we are leveraging our expertise in microelectronics to advance a number of important initiatives at the QSC, while simultaneously leveraging expertise in the QSC from outside of HEP to improve searches for dark matter and other new physics powered by quantum sensors.
\par 
Early access to quantum computers is needed to develop a deeper understanding of how these novel means for computation can be used efficiently and scalably for HEP problem sets. This includes opportunities to test and evaluate existing quantum technologies as well as efforts to tailor future iterations of these technologies to address computational bottlenecks. These efforts will require new tools and testbeds that supports research by the HEP community focused on key challenges in quantum simulation, data analysis, and many other areas. In particular, a deeper understanding of how encoded models for quantum physical processes can be efficiently simulated is essential for demonstrating a quantum computational advantage.
\par 
Therefore, timely research to enable efficient deployment of error-correcting codes in HEP-relevant calculations is needed.
The highly successful campaign of the quantum chemistry community over the past decade serves as an example of how to incorporate available error correction methods into quantum computing applications of interest.
Here, research has systematically driven down required circuit depth and numbers of qubits by  orders of magnitude.
Although we may leverage many of their discoveries, the HEP and nuclear physics (NP) applications of interest are often set in higher dimensional spaces and require quantum gates that effectively operate on larger numbers of qubits.
Circuits involving only two-qubit gates may be prohibitively deep, such that quantifying trade-offs associated with higher-order gates, qudit representations, and the application of error-correcting codes will be paramount.
The theory of error correction for qudits is less developed than for qubit systems, and HEP scientists are anticipating making important contributions in this area, for example, through the SQMS activity~\cite{sqmswhitepaper}.
\par
Research and development of individual quantum computers, including their methods, tools, and techniques, should be augmented by the study of how such interconnected devices and systems operate. Scaling up quantum computing technologies will require integrating components into networks that function in concert through distributed and synchronized operations. Quantum networks offer the benefits of entanglement expanded across space and time scales, which expose new opportunities for data processing as well as communications and sensing.
This includes, for example, the concepts of distributed quantum clocks for precision timekeeping as well as entangled sensors for enhanced measurement sensitivity.
The integration of these specialized quantum networks with quantum computing will require unique components for transducing quantum information between different physical media as well as methods for storing and correcting such information.
To enable these advances, it will be crucial to develop means to extend the quantum channel distance (e.g., quantum repeaters).
HEP competencies in controls, fast timing (synchronization), and system integration are essential for developing functional quantum networks.

Workforce development is an area where the HEP quantum computing program is expected to have significant impact.
HEP has long had an outsized role in national workforce development, producing a large number of highly-trained, technologically capable, and analytically minded individuals that have made impacts throughout the economy.
This is because the HEP science mission brings powerful tools and technologies to bear on the most fundamental questions about nature.
This has always attracted budding  scientists.
Engaging with quantum computers is strongly centered in this tradition, and the ability of HEP science to both benefit from, and make useful contributions to QIS provides HEP with an opportunity to grow our already significant footprint in education and workforce development.

It is crucial to empower current HEP scientists with QIS tools for their own immediate research problems and as a platform for training junior scientists.
The workforce development plan should ensure that HEP scientists engaged in quantum computing research have the opportunity of full career paths within HEP.
This will aid in retention and recruitment in this highly competitive area.
It is crucial to establish multiple ``critical mass'' activities in this area to enable long-term success, and identifying ``homegrown'' solutions by promoting and retaining from within HEP is important.
There must be support and positions at universities to enable a robust higher-education pipeline, undergraduate and graduate courses in physics departments, and active collaborations with engineering, computer science, applied math, and statistics departments as well.

Much of the QIS-HEP activity has been supported by the Department of Energy (DOE), Office of Science, QuantISED program.
Launched in 2018, it followed from a series of community  studies and discussions. It has relied upon interdisciplinary collaborations between HEP and QIS researchers and partnerships between national laboratories, universities, and hardware vendors such as IBM, Rigetti, and Google. Major topic areas addressed by QuantISED include: cosmos and qubits, foundational theory, quantum simulation, quantum computing, quantum sensors, and quantum technology.
The creation of the SQMS Center at FNAL as one of the NQI centers is providing a further significant source of support for research at the QIS-HEP interface.
It is important to grow and stabilize these efforts with a robust QIS ``base program'' in HEP.
Stable support will empower long-term planning and enhance program flexibility and agility in the face of a rapidly changing quantum technology landscape.

It is also important to recognize the role small and modest seed grants have played in developing QIS capabilities in HEP.
Lab Directed Research and Development (LDRD) programs have been a fertile ground for initiating new projects that have grown into significant efforts.
Along with a healthy supply of large, ambitiously funded projects, it is also necessary to provide small and starter funding to give new ideas an opportunity to gain traction.

Building upon the progress of recent years, the upcoming period will be a formative time for the integration of HEP and QIS.  Important pathways and best practices will be established, including low-barrier-to-entry, robust and reliable access to quantum computers, sensing capabilities and communication (at the speed of science) for a diverse profile of HEP researchers, analogously to the successful classical computing programs.
Many problems of interest in HEP ultimately require quantum computing capabilities to achieve our science goals.

\newpage

%%%%%%%%%%%%%%%%%%%%%%%
\section{HEP for QIS and QIS for HEP}
\label{sec:intro}
\noindent
Quantum computing, communication, and sensing
promise capabilities that lie far beyond what can be achieved classically through the manipulation of quantum correlations, entanglement and coherence - defining the second impact era of quantum mechanics.
The inherent non-locality of entanglement, once considered a problematic aspect of quantum theory, is  enabling technologies 
that will fundamentally change 
how we acquire, transmit, process, encrypt and store information. 
The remarkable progress to date has been brought about by multi-disciplinary  collaborative research at universities, national laboratories and technology companies. 
HEP continues to play a critical role in establishing and elucidating fundamental aspects of quantum information, and  
driving ultra-sensitive quantum sensing to search for new particles and new laws of nature, 
important for QIS and  applications outside of HEP.
During the coming decade and beyond, 
it has a unique role to play in advancing quantum information science and technology (QIS), and will be transformed by the rapid advances in QIS that are being brought about by an evolving  ecosystem of targeted research, engineering and technology transfer.
During the coming decade, QIS-related research within HEP is expected to become a major and foundational component of the HEP mission.

The discovery of quantum mechanics in the early 
$20^{\rm th}$ Century
ignited advances in science and technology that had been unimaginable.
Combined with special relativity, 
fundamental research into quantum  field theories and the basic building blocks of the universe 
defined the high-energy physics research mission, and led to the establishment of the Standard Model as the robust description of everyday matter. 
The significant efforts to establish a more fundamental origin of the Standard Model and unification with general relativity, naturally lead, in part, to a convergence with research around fundamental questions about the nature of information and computation.
Within the HEP community, since the early 1980's spurred by Feynman's efforts to understand computation at a fundamental level and the limitations of classical computing,
HEP has spearheaded theoretical advances in understanding entanglement in a range of quantum field theories, 
forming the basis for major broad-based efforts to develop quantum error correction protocols, 
starting in the mid-1990s.

The last few years have witnessed a sea-change in 
QIS-related computational activities within HEP.
The availability of cloud-accessible quantum devices at technology companies, such as IBM, D-Wave and Rigetti, and increasing user-friendly access portals at Microsoft, Amazon and others,
combined with an environment that encourages and promotes collaboration between HEP researchers, software developers and engineers at technology companies and national laboratories,
kick-started the HEP community's efforts.
While a few years behind quantum chemistry in embracing QIS, HEP and other domain sciences such as Nuclear Physics (NP), have started a coordinated program to assess quantum resource requirements to address HEP scientific objectives.
The increasingly diverse array of digital, analog and hybrid quantum devices, with qubits, qudits or continuous variables, controlled in systems of  cold-atoms, trapped-ions, and superconducting qubits, optical, and superconducting radio frequency (SRF) cavities, and quantum annealers, with distinctive
capabilities, ever increasingly sophisticated quantum algorithms and software,
are providing a rich environment to assess utility for HEP challenges.  
It is currently too early to down-select architectures and systems for HEP specific objectives, and the next period will be important to inform such possible future decisions.

The NISQ-era is defined by devices that are noisy and intermediate-scale that have some degree of noise mitigation but without error correction.
Algorithms and workflows are becoming more sophisticated in noise mitigation, and only recently have elements required for error-correction been demonstrated on quantum devices.
While it is anticipated that a quantum advantage for HEP processes could be found during the NISQ-era, error-correction will be required for precision calculations at scale.
Advancing quantum devices to address HEP challenges is just one aspect of the importance of advancing quantum simulations and computing.  
While much effort is placed on isolating a device from the environment to preserve quantum correlations for computation, quantum sensor wish to optimally couple to certain aspects of the environment, with quantum algorithms processing acquired signals.
The complementarity of these systems is evident.
Somewhat similarly, improvements in analog quantum devices for simulation impacts the use of analogous devices for detecting acceleration, important for positioning devices that could be useful when, for example, GPS is unavailable.

The HEP community has launched a ``design and build'' initiative in quantum computing that utilizes its expertise and technological capabilities developed for particle accelerators.
FNAL is the lead institution of the SQMS NQI Center that is one of five such centers supported by DOE.
It is focused on advancing and materials for SRF cavities to build systems that can perform quantum computations for, for example, quantum field theories.  This is a major activity for the nations QIS program and should be considered a large scale activity within the Snowmass process.

While not the main focus of this report, it is the case that HEP activities impact future quantum communications and the quantum internet.
Quantum communication is enabled by transmission of entangle quanta, quantum repeaters, transduction and general quantum technologies that all involve quantum manipulations required for quantum simulation and computing.
In addition to the direct benefit of quantum simulation to HEP, these advances are also expected to impact quantum communications technologies.
Furthermore, quantum networks appear likely to be a necessary component for scaling quantum computers to a sufficient size (in terms of number of qubits) to effectively tackle HEP science problems.

This report should be considered in the broader context of the Snowmass activity.  The quantum activity within HEP is distributed among a number of areas that will also provide reports related to quantum, including the Theory, Accelerator and the Instrumentation, Underground Facilities, Rare Processes and Precision Frontiers.
In this light, we have not attempted to be comprehensive, and have restricted ourselves to topics and issues directly related to computation, and have relied heavily on the Whitepapers contributed by the community that can be found on the {\tt arXiv}.

%%%%%%%%%%%%%%%%%%%%%%%
\subsection{HEP and the national quantum ecosystem}
\label{sec:hepnatecosystem}

It is valuable to prove the worth of HEP's engagement with QIS to the quantum community because quantum technology and research is potentially crucial for our science, making partnerships deeply valuable.
But it is also the case that QIS represents a ``growth opportunity'' for HEP.
Quantum technologies are poised to become a very large part of the economy, 
potentially rivaling the "silicon economy" in the future,
and research investments in QIS are similarly primed for rapid growth to fully enable the benefits in a competitive international landscape.
By clearly establishing HEP as an important early contributor to quantum research, we can expect to benefit in terms of long-term investment in the HEP science program.

Quantum computing is like many nascent technologies --- it is caught between a need to deliver results to justify investment, while also requiring fresh investments to produce results.
It is widely believed quantum computing offers long-term broad value with a rich set of business and scientific applications, but sustaining the levels of investment required to ultimately deliver these results is a challenge because there are so few near-term applications of practical impact.
In this sense, one of the most important contributions HEP can make to QIS is to be an early adopter of quantum computing.
While many businesses would like to eventually utilize quantum speed-ups for applications, they are uninterested in slower versions of the machines. 
However, in HEP, even NISQ-era quantum computers offer exciting possibilities as simulation platforms and research tools to better understand quantum physics.
What looks like a ``bug'' to others is a ``feature'' for us.
Furthermore, many scientific applications stand to benefit from quantum advantages long before commercial applications, which makes our engagement substantially easier to justify.

HEP brings a variety of useful technical and managerial competencies to the national quantum ecosystem.
Perhaps the most prominent example is our long history with world-class superconducting devices.
Our deep expertise in materials science and device physics for superconducting cavities, developed over decades of innovation in the accelerator program, offers very interesting new capabilities to quantum information processing.
For example, using superconducting cavities as part of a quantum computer was pioneered at the prototype stage in the QIS community, but HEP brings the capability to produce cavities with quality factors orders of magnitude larger to the effort, as will be discussed further in Section \ref{sec:sqms}.
We also have deep expertise in fast and custom electronics for large and complex detectors, and extensive experience in cryogenic electronics --- all of which are crucial capabilities for scaling approaches to quantum computing that utilize superconducting technology.
It is also common in HEP to deal with very large projects and collaborations, and we have experience scaling up some of the most complex systems ever built.
This expertise translates naturally into QIS, especially in these early days where science and research applications are the main focus.

Finally, it is worth noting the international nature of HEP provides certain advantages in terms of contributing to the QIS landscape.
In particular, our deep ties with CERN and other international laboratories allow us to tap a network of expertise that will accelerate applications.
For example, scientists at CERN are keenly interested in using quantum algorithms for HEP data analysis and are investing intellectual effort in that space.
This will inevitably bear interesting fruits and through the interactions of HEP science, these advances will come back to QIS and quantum computing more broadly.

\subsection{The role of HEP in developing a QIS-enabled workforce}
\label{sec:workforce}

HEP has long played an important role in the national STEM education pipeline and ecosystem.
Because HEP science requires the relentless advancement of many technologies across a multitude of domains, and because the fundamental nature of its program of inquiry is so appealing to the seeker in the human heart, we have consistently attracted brilliant minds to the field, eager to meet the challenges and answer the questions of HEP.

Our engagement in QIS only grows this portfolio of inquiry, and opens exciting new challenges on the ``entanglement frontier,''
This means we are already well-situated to play a key role in preparing a quantum-ready workforce.
We not only provide an environment where direct QIS research and training are possible, but also an exciting avenue for cross-disciplinary and applied QIS work.
Quantum computing is not quite yet ready to solve logistics or materials science questions, but the workforce that will services those industries can be trained working in HEP today where our science program is already poised to benefit from QIS.

Workforce development and training are important areas for HEP scientists to invest effort and resources.
We should ``cross-polinate'' with QIS summer schools and internship programs and look for more opportunities to tap into the growing swell of interest in this area.
HEP has already leveraged opportunities like the Department of Energy's Visiting Faculty Program, but building new internship and workforce development efforts provides an important opportunity to redress historical failings in properly engaging students from minoritized backgrounds.
It is imperative that new efforts actively seek to support more women and under-represented, minoritized students in HEP.

It is also important to plan support for tenure-track university hires in physics departments, and at national laboratories.
HEP science programs have benefited enormously from postdoctoral fellows and other young scientists and engineers that have moved into HEP from pure QIS backgrounds.
These individuals need to be able to find permanent homes in HEP --- both because we cannot afford to lose the investments we have made in their training or their accumulated expertise and because we must demonstrate a viable career path if we are to continue to be able to hire and recruit in what is an incredibly competitive market for talent.
We need to build the ``critical mass'' of expertise required for a self-sustaining program with the capacity to grow its own QIS talent, and it will likely require planned and dedicated support to nucleate these programs across the country.

%%%%%%%%%%%%%%%%%%%%%%%
\subsection{This is a pivotal moment}
\label{sec:pivotalmoment}

We are at a crucial point in the history of quantum computing.
Quantum volume \cite{quantumvolume}, one of the most widely used metrics for the problem complexity accessible by a quantum computer, has entered an era of exponential growth \cite{ibmqv}.
Error corrected quantum computers are on the public roadmaps of some of the major companies working in quantum computing, with delivery anticipated as soon as within the next decade \cite{googroadmap,ibmroadmap}.
Quantum error correction (QEC) is almost certainly a pre-requisite for, and enabler of, every serious practical application of quantum computing that is beyond the reach of classical computers, and the expected impacts are potentially civilization-changing in scale.
While it isn't clear yet exactly which will be the first beyond-classical applications of quantum computers as we transition into the era of fault tolerance, from quantum simulation alone we can expect tremendous advances in materials science, drug design and pharmaceuticals, and chemical engineering, to say nothing of fundamental advances in understanding spin systems and field theories.

This is the first Snowmass to prioritize quantum information --- our choices will be very influential in terms of establishing QIS within HEP and solidifying HEP's position to justify a claim on the rapidly growing investments in QIS on the national stage.
Furthermore, we have the opportunity to seriously consider physics grand challenges that were beyond our reach before in lattice gauge theory and the fundamental physics of information.
HEP has important responsibilities and  opportunities with respect to our core science mission and advancing QIS at the national level.
It is important to engage at the most ambitious level possible.

%%%%%%%%%%%%%%%%%%%%%%%
\section{HEP and the National Quantum Information Act}
\label{sec:nqia}

The National Quantum Information Act (NQIA) was an  legislative effort in 2018 to prepare the United States for leadership in a post-quantum computing world.
Among many provisions, the NQIA established a multiple well-funded national research centers designed to serve as hubs for innovation and scientific advancement in the field of QIS. A notable example is the Superconducting Quantum Materials and Systems (SQMS) Center \cite{sqmswhitepaper} led by Fermilab, the Department of Energy's (DOE's) only single-purpose National Laboratory with a mission focused on HEP. 
Establishing SQMS at Fermilab as part of NQIA was an endorsement from the highest level that QIS and HEP science are intertwined and crucial for each other's long-term success.
Additionally, there is a strong HEP presence within the other DOE National Quantum Information Science Research Centers. This includes, for example, the Quantum Science Center (QSC) hosted by Oak Ridge National Laboratory (ORNL) and its focus on applications of quantum computing for both high and low-energy physics as well as many other scientific domains.
ORNL is a large, multi-purpose laboratory and the inclusion of an HEP research agenda within QSC is strong evidence that QIS benefits from collaborative engagements with multiple disciplines.
More broadly, the national QIS ecosystem is bringing together stakeholders from across scientific domains to address common concerns and shared priorities for QIS research.

\subsection{The Superconducting Quantum Materials and Systems Center}
\label{sec:sqms}
\noindent
{\it Note : The contents of this section are, in part, taken and modified from {\bf Quantum computing hardware for HEP algorithms and sensing}~\cite{sqmswhitepaper} }
\vskip 0.1in

The primary goal of the SQMS Center is to understand and mitigate quantum decoherence, and to deploy superior quantum systems to advance applications in quantum algorithms and sensing.
The SQMS Center (colloquially, ``SQMS'') is discussed in detail in a Snowmass Whitepaper.
Please see \cite{sqmswhitepaper} and the extensive references within.

\begin{figure}
  \centering
  \includegraphics[height=0.2\textheight]{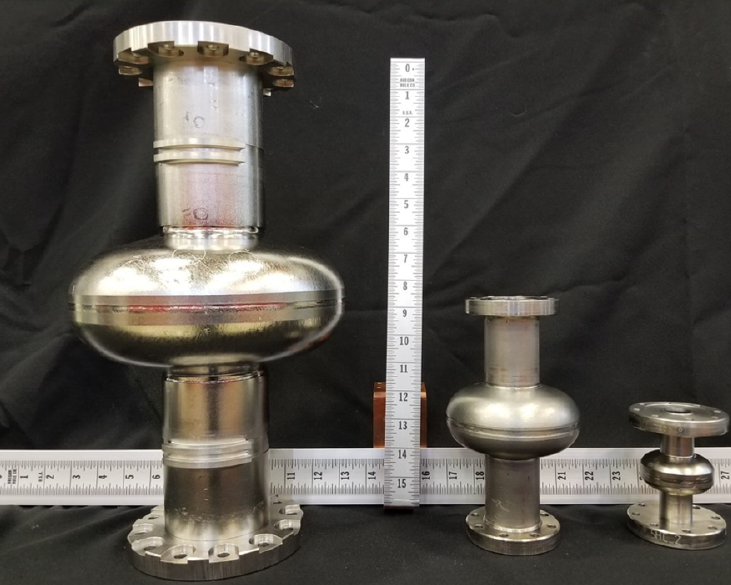}
  \caption{Single-cell TESLA shape SRF cavities.}
  \label{fig:singlecellcavity}
\end{figure}

At SQMS, the technology and expertise developed by the HEP community, primarily based on our needs in particle accelerators, provides exceptional theoretical and experimental resources to advance the physics of decoherence.
We are able to construct cavity oscillators with the highest $Q$-factor in the world --- the highest of any synthetic device, in fact.
See Figure \ref{fig:singlecellcavity}.
This is a crucial contribution that HEP is extremely well-suited to make to the national quantum ecosystem and a prime example of the mission of HEP benefiting from engaging with QIS and the national quantum ecosystem benefiting from engaing with HEP.
The oscillators, when coupled to a qubit, create a powerful new quantum information processing platform with enormous potential for impact specifically on HEP science applications but also more broadly.

Three-level (qutrit) and $N$-level (qudit) systems more generally may offer important advantages for HEP scientific applications that require very-high qubit-count operators.
These operators are challenging to construct from two-qubit operations --- they produce extremely deep circuits.
The advantages of going even to just three-levels are significant and may actually enable NIQS-era applications of scientific value. 
Looking forward, quantum information processors built using superconducting radio-frequency (SRF) cavities may be extremely well-suited to implementing the high-dimension operators required by HEP science due to the intrinsic all-to-all connectivity.

The extremely long coherence time offered by these cavities motivates a number of QIS applications, but we must introduce a non-linear element (e.g., a coupled transmon qubit) for universal quantum control.
This in turn motivates a research program built around minimizing associated losses and designing optimal control schemes.
It also motivates research around qubit technologies that may be less disruptive to cavity coherence, such as fluxmon qubits, or $\cos \phi$ qubits.
Fully utilizing cavity systems for HEP applications will require an in-depth understanding of the different pros and cons of the various encoding schemes available.
Fock-basis encoding is the current leading, and best studied, contender, but it may not prove the most optimal in the end when quantum error correction is properly included.
For example cat-codes, in which the information is encoded in a basis spanned by cat-states, offer some natural advantages when considering errors due to photon losses (which are expected to be the dominant error mode).
Gate engineering more broadly is an important area of research for 3D cavity QPUs.
Early work suggests universal control is achievable, but special emphasis needs to be placed on making sure our methods for gate synthesis are scalable to the system sizes we would like to construct.

For quantum sensing applications, again superconducting cavities may offer tremendous advantages as probes for fundamental physics due to the extremely long coherence times these devices offer in the quantum regime (relaxation times are about 2 seconds --- internal quality factors $Q_0 > 10^{11}$).

Interconnecting different multi-mode cavities, with each controlled by a single qubit, is an important scaling strategy to investigate and develop.
SQMS is exploring the use of multi-cell cavities as multi-mode quantum systems, and methods for engineering generalized gates that may operate on the combined Hilbert space of the multi-cell system.
The challenges in the implementation of quantum networks through interconnected nodes and hybrid transduction devices is a crucial area for inquiry.
Beyond distributed computation, applications to networks of quantum sensors is of particular interest in HEP, as these devices may enable measurement sensitivities below the standard quantum limit (SQL).
HEP, especially through the SQMS, has an enormous amount to offer to these challenges, and the potential to benefit a great deal from their resolution.
Quantum transduction technologies that exploit 3D bulk niobium cavities are particularly promising and could be critical technology far beyond HEP.

Materials research is another crucial area of investment to deliver 3D QPUs capable of making progress on the quantum computing applications of the greatest importance to HEP.
The incredible potential of these devices has been enabled by decades of research and development in the accelerator community, and this area a prime example of the capacity for HEP to make crucial contributions to QIS on the national scale by leveraging our world-leading expertise in superconducting devices.
A variety of purification techniques have been developed to improve the quality of niobium cavities --- as previously mentioned, these cavities recently demonstrated photon lifetimes of up to 2 seconds.
Further improvements will be driven by deeper understanding of impurities in niobium.
Furthermore, HEP materials research will drive improvements in the coherence times of other qubit technologies as we come to understand two-level systems (TLS) defects better.

Another area where HEP expertise can lend real strength to the national quantum ecosystem is in both cryogenic and room-temperature micro-electronics.
Because HEP has been driven by the stringent requirements of triggering in particle physics experiments, as a field we have developed highly specialized capabilities in areas that have become of very broad importance.
This offers opportunities to further grow this expertise for applications within HEP by leveraging our impact in disciplines outside traditional particle physics.

Quantum error correction is of critical importance to quantum computing.
Beyond algorithmic innovation, there is substantial room to develop qubit technologies with improved coherence times.
And for cavity-based computation, investments in theory-based error correcting codes and feedback systems may offer tremendous returns for applications of interest in fundamental physics.

Proper access to quantum computing is a major challenge and prerequisite for an empowered community.
The HEPCloud project aims to provide the user community with simple access to quantum devices, removing many of the headaches of interfacing with the hardware directly, or even with a variety of cloud vendors.
There are important questions to keep in mind when designing analysis, storage, or other advanced computing facilities.
In many cases quantum computing will be simple to integrate but in other cases it will require substantially more thought.
When designing the computing facilities of the future it will be important to remain flexible and supportive of the integration of quantum computing.

\subsection{The National QIS Ecosystem}

Within the US, the NQI Act has served to bolster a network of partnerships between QIS stakeholders including  23 federal agencies and more than 13 major research centers. This national QIS ecosystem offers a backbone upon which the field will continue to develop, and it is important for the HEP community to remain engaged fully in the broader partnership.
This engagement will ensure collaboration in addressing common concerns while advancing  state-of-the-art research in QIS.
\par
Central to the QIS ecosystem are the series of major NQI research centers and institute sponsored by the Department of Energy (DOE), the National Science Foundation (NSF), the Department of Defense (DOD), and the National Institute for Standards and Technology (NIST) among other federal agencies. These research programs address a multitude of technical and programmatic concerns that facilitate the overall growth of the QIS ecosystem. While each approach is tailored to the agency mission and core capabilities, the collection of these efforts represents a comprehensive, whole-of-government effort. 
\par
For example, the National Quantum Information Science Research Centers (NQISRC's) supported by the DOE serve as focal points for the integration of QIS skills with national laboratory facilities. The NQISRC's are advancing the development and readiness of quantum science and technology by fostering innovation, workforce development, and technology transfer as well as cutting-edge research. While the SQMS center, highlighted above, aligns clearly with the HEP research program outlined in this document, it is imperative to recognize that each of the NQISRC's contribute to advances in quantum computing that are essential for the success of QIS research. The Quantum Science Center at ORNL is developing new materials for topological quantum computing as well as methods of quantum sensing applicable to searches for axion dark matter candidates, while the Codesign Center for Quantum Advantage hosted by Brookhaven National Laboratory is building new quantum computers using superconducting microwave circuits for scientific applications. Q-Next at Argonne National Laboratory and the Quantum Systems Accelerator at Lawrence Berkeley National Laboratory each offer novel approaches to quantum simulation testbeds using variety of quantum technologies. 

Similarly, the NSF has sponsored over a dozen research centers or institutes across the US to spur research as well as education in QIS. 
For example, the Quantum Leap Challenge Institutes (QLCI) have funded five major centers to emphasize development in sensing and measurement, hybrid quantum architectures and networks, scaling quantum computers, improving quantum simulation, and in sensing in biophysics and bioengineering.
Additionally, the NSF is using programs like the Quantum Computing and Information Science Faculty Fellows (QCIS-FF) to expand the academic workforce and build a strong community of QIS scientists.
More information on these programs and others may be found at the NSF \footnote{\protect\url{https://www.nsf.gov/mps/quantum/quantum_research_at_nsf.jsp}}.

%%%%%%%%%%%%%%%%%%%%%%%

%%%%%%%%%%%%%%%%%%%%%%%
\section{Access to quantum and classical computing resources}
\label{sec:accesstoresources}
\par 
A fundamental concern for advancing the research and development of quantum computing methods is access to these quantum computing platforms. Access to quantum computing platforms is essential for developing, testing, and evaluating applications in simulation, machine learning, data analysis, particle tracking, and many other tasks~\cite{humble2022advancinghep, opensystemsim}. Access enables validation of the underlying quantum algorithms and the supported applications as well as the verification that the developed hardware platforms are both accurate and efficient.
\par 
While many different facets of science and industry would benefit from access to quantum computing, there are specific needs for HEP that warrant specialized models of access. Application-driven co-design of quantum computing system as well as problem-specific error correction and error mitigation strategies are likely necessary to optimize the quantum computational advantage expected within the HEP context. That is to say, addressing scientific grand challenges that could not be answered using only conventional resources will require a dedicated and focused effort, and the HEP community has identified several rallying points for such efforts.
\par 
For example, it is important to understand whether optimal tuning of quantum computing hardware, including the layout and connectivity, gate duration and pulse design, and noise tolerance, depends intimately on program execution and purpose. For both small and large-scale systems, the performance difference between general-purpose quantum computing systems and special-purpose, dedicated simulators may be significant and the topic is worth further investigation. As discussed above, the SQMS Center hosted by Fermilab is pursuing a quantum computing concept based on qudit representations that may offer specialized advantages for solving field theory and many-body quantum problems for precisely these reasons. Similarly, recent research has argued that tailored hardware design, e.g., qutrits in favor of qubits, may offer particular advantages for encoding problems in HEP~\cite{Bauer:2022hpo}. 
These efforts underscore the expected benefits of specialized encodings benefit both in and beyond the HEP problem set, e.g., there is a strong theoretical overlap between the field-theoretic and materials models that may make these specializations broadly supported. 
Similar dedicated efforts echoed by other research centers and institutions will be important for the HEP community to prioritize access as a means of testing and validation.
\par
Access to the rapidly evolving advances in quantum computing hardware will benefit the specialization of quantum computing to HEP problem sets. As noted above, there a multiple technical consideration related to the encoding and processing of models and data that may warrant specializations. Evaluation of these benefits will be made possible through collaboration design of the hardware for such applications. Robust participation in quantum computing testbed programs will help HEP answer these questions about whether specialized or dedicated hardware is warranted. For example, these activities would include investigation of how to best execute low-level control to achieve application goals. More broadly, hardware-software co-design efforts that engage domain scientists with specialists in computer science and math have been a proven strategy for success in conventional computing and we expect this will remain true in quantum computing. Cutting-edge domain applications help motivate and drive fundamental progress in computer science and math that will have spill-over effects into other fields. Similarly, by remaining engaged in cross-disciplinary projects, the HEP community can benefit more quickly and efficiently from advances originating in other areas.
\par
Continued access to quantum computing systems is also necessary for reliable and reproducible results by the community of users. Presently, quantum computers demonstrate transient noise processes that induce drift in the calibration of quantum computer control systems and uncertainty in the device behavior. Access for testing and validation noisy devices requires methods to mitigate errors and performance reproducible demonstrations. This current paradigm of validating quantum computing applications using noisy, small-scale quantum devices is likely to evolve quickly over the next decade toward larger systems with more reliable behaviors. Ultimately, quantum computers may demonstrate fault-tolerant quantum-error corrected operations, but during this transition period, access to quantum computing systems will be critical for maintaining verification of the algorithmic designs as well as performance validation of the burgeoning applications. This is underscored further by the anticipated divergence between quantum and conventional computing capabilities, which make the latter computers insufficient for future validation of the `entanglement frontier'.
\par 
Currently, there are several means by which the research community may access quantum computing systems. Some commercial vendors provide access to quantum computing devices using remote-access models. This includes companies such as IBM, Rigetti, IonQ, D-Wave, and Quantinumm among others. Still others partner with aggregation services like Azure Quantum from Microsoft or Braket from AWS that re-sell and manage access to multiple vendors. In addition, national research agencies offer support for access through their facilities. This includes the US Department of Energy Quantum Computing User Program and the Testbed program as well as the ABAQUS program at CERN. Finally, many research institutions support in-house experimental laboratories through which collaborators may engage in testing and validation, for example HEPCloud at Fermilab \cite{opensystemsim}.
\par
Alongside access to quantum computing resources is the accompanying integration of these resources with quantum networks, as detailed in Section~\ref{sec:quantumnetworks}. Quantum networks have the potential to integrate many different technologies, including time and frequency distribution networks, quantum data technologies, quantum networking, and quantum computing, that serve fundamental and transformative role for high-energy physics experiments in  different energy regimes. For example, these networks can provide standard reference signals for anti-matter and low-energy experiments, such as ASACUSA and ALPHA, using high-precision laser spectroscopy. Notably, HEP-driven adoption of integrated quantum technologies will require community-led developments to acknowledge existing practice as well as standardized interfaces and methods.
\par 
Access to conventional high-performance computing (HPC) resources has played an essential role in the study of high-energy physics, and this is likely to continue for the next decade as well. Now, however, quantum computing resources may complement these HPC resources as either specialized accelerators or as alternative platforms to advance computational studies. The accompanying development of quantum computing tools, described further in Section \ref{sec:quantumtoolsandtestbeds}, must then also be considered within the context of HPC workflows used for HEP problem sets.
Quantum computing should not over time require separate computing infrastructures, application platforms, or languages, but must be integrated into the existing cloud-based physical and software infrastructure. This will enable the deployment of hybrid, heterogeneous workloads where the quantum processing units (QPUs) are part of a broad range of possible accelerators. Memory requirements scale steeply with every qubit added in a numerical simulation and this quickly reaches the limits of convention systems. Consequently, quantum simulations of large scale systems may be performed by means of distributed and parallelised computation across HPC clusters of CPU/GPU \cite{opensystemsim}. Novel architectural frameworks to integrate quantum capabilities APIs in cloud architectures are also emerging.

%%%%%%%%%%%%%%%%%%%%%%%
\section{Quantum Simulation for Fundamental Science in HEP
} 
\label{sec:quantumsim}
\noindent
{\it Note : The contents of this section are, in part, taken and modified from {\bf Quantum Simulation for High Energy Physics}~\cite{Bauer:2022hpo} }
\vskip 0.1in
Research and discovery in HEP has flourished, in part, from the unique capabilities of computation to process information and derive reliable and robust predictions from theory that cannot be obtained by other means.
High Performance Computing (HPC) continues to evolve, enable and accelerate progress in a broad range of  HEP sub-disciplines.
As a result, HEP physicists are in a prime position to take advantage of the potential benefits of rapidly-advancing quantum-simulation hardware and its growing algorithm/software/compiler/user ecosystem. They have, indeed, identified a range of problems outside the capabilities of emerging Exascale classical computers and beyond, spanning subfields such as collider physics, neutrino (astro)physics, cosmology and early-universe physics, and quantum gravity.
The simulations of relevance to HEP involve quantum field theories at their core. 

Theoretical efforts have gained intensity in recent years to frame quantum field theories in the language of Hamiltonian simulation, which is the natural language of quantum simulation. One important element of these efforts is the development of protocols to enable a complete quantification of uncertainties, including those from digitization of continuous fields onto discrete Hilbert spaces.
While algorithmic progress for Abelian and non-Abelian gauge theories has been significant in recent years,  efficient algorithms are likely not going to be the early ones. 
Furthermore, the asymptotic scaling with simulation accuracy 
is expected to be significantly improved in coming years by appealing to physics guidance and empirical analyses.
Algorithms necessary for preparation of initial hadronic  states to simulate scattering, structure functions, and much more,  need to be  advanced. 
Nascent progress in reliably defining entanglement in quantum lattice simulations will continue.

Theoretical, algorithmic and hardware developments need to continue in a coordinated way.
For instance, algorithms for state preparation, time evolution, and measurement and tomography 
have been dramatically 
improving within the QIS community. It is not yet known what the best theoretical formulations of 
HEP problems are coupled with these algorithms, or if new algorithms tailored to field-theory simulations are needed. 
Such algorithms need to be concrete, allow for realistic estimates of quantum resources required to execute them given a desired accuracy, and take advantage of physics input such as locality, symmetries, and gauge and scale invariances. 
To operate simulations in an analog or hybrid analog-digital mode, and engineer quantum field theories of increasing complexity and relevance, enhanced modalities and advanced quantum-control capabilities must be developed and added to current state-of-the-art platforms, 
and a range of physical architectures from atomic, molecular, optical, SRF-cavity and solid-state systems should be carefully evaluated.
To ensure the feasibility of the simulation approaches, prevent a wide gap between theorists’ proposals and experimental realities, 
and tighten the theoretical algorithmic scalings by supplementing empirical observations, implementation and benchmarking using NISQ hardware will continue to be an essential endeavor for the HEP community. 
General-purpose error-correction and noise-mitigation methods, or those tailored to HEP simulations or inspired by HEP developments, are a necessity, as are hybrid classical-quantum algorithms with clear cost-saving benefits. Furthermore, to enable the broader HEP community to readily participate, models for programming quantum computers need to mature to a level in which abstractions and library-based methods can be adopted to expedite programming and ensure portability. Efficient compilers that can take the programs written in high-level programming languages to efficient low-level code are needed for HEP applications. 
The quantum-computing paradigm likely requires that users effectively engage across the quantum-computing stack, 
and have access to software layers that form and 
sequence native device controls specific for interactions of relevance to HEP applications.

Since quantum technology is advancing rapidly in universities, national laboratories, 
technology companies and start-ups, it is important that collaboration and cooperation is ensured between all these sectors. A  study to determine whether industry-developed quantum hardware satisfies the needs of the HEP community is required.  
HEP likely needs to play a central role in the development of quantum simulators, while at the same time needs to have access to those platforms developed in industry and other partners. Such multidisciplinary research requires training, retention, and empowering the workforce, and in some cases, retooling and reorienting the talent in HEP to enter this rapidly advancing field. Importantly, diversity and inclusivity will be crucial in ensuring an intellectually open, sustainable and strong program at the intersection of HEP and QIS.

%%%%%%%%%%%%%%%%%%%%%%%%%%%%
\subsection{Quantum computing and event generators}
\label{sec:eventgenerators}
\noindent 
{\it Note : The contents of this section are, in part, taken and modified from the Snowmass Whitepaper: {\bf Quantum Simulation for High Energy Physics}~\cite{Bauer:2022hpo} }
\vskip 0.1in
Particle collisions are used to look for deviations from the SM predictions, which are expected at some scale due to the inability of the SM to describe well-established facts about nature, such as the existence of dark matter and the matter-antimatter asymmetry.
They are notoriously difficult to describe theoretically, largely due to the fact that they are governed by physics over widely different length scales. 
While particle collisions are often aimed at discovering processes happening at the highest energies (shortest distances), the observed distribution of final states is affected by physics at energies ranging from the large kinetic energy in the colliding quarks and gluons, all the way to low energies describing the binding of partons into hadrons. Such high-energy collisions typically give rise to a large number of final-state particles, making them too complicated to be readily calculable using Feynman diagrams.

For these reasons, particle collisions are nowadays described theoretically using various types of approximations, each valid in a certain energy interval. 
Processes at the highest energies typically involve only a small set of final-state particles, allowing a perturbative evaluation of the full quantum-mechanical amplitudes. 
The production of the large number of additional partons is traditionally described by a parton-shower algorithm based on classical emission probabilities rooted in a collinear approximation in the limit of infinite number of colors. 
Finally, phenomenological models are used to describe
hadronization into  color neutral hadrons. 
The parton-shower algorithm is typically combined with  hadronization models to allow so-called exclusive event generators, which take a partonic state produced in a short-distance process and turns it into a fully exclusive final state containing only stable or long-lived particles which can be observed in a particle detector. 
Much work has been devoted to calculate the short-distance processes as precisely as possible in perturbation theory. 
Care is required to avoid double counting when combining parton-shower algorithms with short-distance calculations at higher orders in perturbation theory. 
While the combination of these ingredients allows for simulation of fully-exclusive scattering processes, the presence of the various approximations imply that the resulting distributions are robust for certain observables, but it is difficult to estimate the uncertainty in the obtained predictions.
In order to reduce the sensitivity to the details of the modeling of hadronization effects and to the approximations made in the parton shower, comparisons of the obtained simulations to experimental measurements are limited to observables that are “sufficiently inclusive”. 
While parton showers allow for the calculation of less inclusive observables, the results are much more dependent on the particular choices made in the modeling, and therefore, typically have significant uncertainties.

Quantum computers hold the promise to simulate (high-energy) scattering processes from first principles, and  without uncontrolled approximations. The basic idea is to discretize the continuous spatial dimensions and to digitize the continuous
field values. 
This turns the infinite-dimensional Hilbert space of the field theory into a finite-dimensional one.
Since the Hilbert space is exponential in the number of lattice points required, it is far too large to allow such a computation using classical computers.
As real-time evolution is known to reside in the BQP-complexity class, it is expected that real-time evolution of lattice gauge-field  theories, 
including scattering amplitudes, 
can be performed on quantum computers, and that with sufficient capable quantum computers, calculations of direct impact upon experiment can be accomplished.
As an alternative to the simulation of the full S-matrix using quantum algorithms, one can also attempt to devise quantum algorithms for the parton shower and the hadronization process. 
The idea of a quantum parton shower is to still work in the collinear approximation that underlies a classical parton shower, but to include quantum interference effects that are not possible in a traditional approach using classical probability distributions. 
For example, a quantum parton shower was shown to be able to include quantum interference effects arising from amplitudes with the same final-state particles, but different intermediate-particle flavors. 
One can also hope to go beyond the large-$N_c$ limit of regular parton showering, and include the quantum interference between different color structures.
Another approach that 
is under evaluation
for collider physics is to simulate the physics at a particular energy scale on a quantum computer, 
while maintaining more traditional approaches for physics at other scales, and employing effective field theories to make connection with experiment, as been successfully accomplished in classical lattice gauge theory.

%%%%%%%%%%%%%%%%%%%%%%%
\section{Tensor networks for the classical simulation of quantum systems}
\label{sec:tensornets}

Simulating quantum field theories using exact methods on high performance resources (e.g., GPUs and TPUs --- with and without quantum noise models) and going beyond $6 \times 6$ qubit lattice arrays is extremely challenging.
$7 \times 7$ is theoretically possible with leadership-class resources, but $8 \times 8$ is impossible with exact simulation and appears to be insufficient to answer all the questions we would like to study even for simple theories such as $\mathbb{Z}_2$.

But what about inexact simulation? 
As long as simulation truncation errors are smaller than the theory errors when going from the lattice to the continuum, we may still do interesting physics.
Tensor networks have been demonstrated to reduce problems with exponential scaling into problems with linear scaling in exchange for accuracy costs that are often relatively tiny.

Tensor networks are objects with multiple indices that may be assembled to represent states of a Hilbert space or transfer matrix, compute operator expectation values, or express partition functions, and they provide an efficient representation of entangled quantum systems and states.
They may be used to reformulate theories on a lattice and map them into other quantum systems --- they then provide methods for classical study or analysis in the mapped representation on a quantum computer.
The material in this section largely follows a Snowmass Whitepaper  \cite{2022arXiv220304902M} on tensor networks for quantum computing and HEP Theory.
Tensor networks have a number of applications in theoretical HEP, and the references within \cite{2022arXiv220304902M} are especially useful for detail.

One particularly important application for tensor networks in the context of this report is the classical simulation of quantum circuits
--- this is a crucial research area for determining the ``cross-over'' points for quantum advantage in scientific applications.
Tensor networks remain a powerful tool for simulating quantum calculations on noisy hardware and may pull applications from the realm of quantum advantage back into the classically tractable given the envelope of ``protection'' provided by theory errors when translating lattice results to the continuum.
They may also play a role in quantum error correction. 
In qubit-based quantum computing, a complete discretization of both field space and spacetime is required, and tensor networks provide a general formulation for achieving this goal.
Certain quantum states that may be represented using a tensor network, like matrix-product states (MPS), may be efficiently prepared on a quantum computer.
Other tensor networks, like projected entangled pair states (PEPS) and the multi-scale entanglement renormalization ansatz (MERA) are important for dealing with relevant physical models.
Tensor networks also enable more efficient state tomography in some cases.

Tensor network methods in lattice field theory, quantum gravity, and QIS have the ability to improve calculations made difficult by sign problems when standard Monte Carlo and sampling methods are deployed because tensor networks themselves do not rely on statistical sampling.
Fully exploiting the power offered by tensor networks requires a significant investment in computer science techniques and co-design engagement with domain specialists in HEP, with the potential for benefiting from research coming from condensed matter physics as well --- tensor networks are resource intensive and are not currently feasible for complex theories in four space-time dimensions, e.g. QCD --- significant theoretical work and software development is required to make tensor networks useful for advancing HEP science questions.
Of particular importance is the interface between classical high performance computing (HPC), which is required for large tensor network calculations, development, and validation, etc., and quantum computing.
    
With regards to method development there are trade-offs between accuracy, computing time, and memory requirements that must be better understood.
Modern computing systems can accommodate one spatial dimension, but large systems with two or more spatial dimensions are beyond current capabilities and we need significant reductions in the size and connectivity of tensors within the network --- much further study is required to simulate 3+1 dimensions.

In terms of software and library requirements, there is a vibrant software ecosystem \footnote{https://tensornetwork.org/software/}, but tasks like determining the optimal contraction order require research --- the task is considered NP-hard and more work is required to understand acceptable heuristics.
Further work is also needed to utilize distributed memory and take various problem symmetries into account.
Coordinated research plans for software development will help ensure the needs particular to HEP are efficiently addressed.

Of course, tensor networks research requires access to large-scale classical computers and HPC facilities --- memory management and I/O bandwidth for very large tensors in memory are critical bottlenecks for efficient computation.
Progress requires a collaborative workforce that spans HEP and ASCR domains, along with other domain science areas like condensed matter physics.

%%%%%%%%%%%%%%%%%%%%%%%
\section{HEP science and quantum error correction}
\noindent
One the major challenges facing the development of robust and reliable quantum computers, and central to developing robust predictions within HEP from quantum simulations, is Quantum Error Correction (QEC) \cite{Bauer:2022hpo,humble2022advancinghep}.
The capability to detect and correct errors throughout quantum computation is far more challenging than the comparable task in classical computing dues to the fragility of quantum coherence and entanglement, which are easily disturbed through interactions with the environment or imperfect classical controls.  
Developing QEC quantum hardware is front and center in the development plans of the technology companies, such as IBM, Google, and Microsoft. They expect to have QEC devices available within the coming decade, which will lead to a further paradigm shift in computing.  Important elements of QEC have recently been demonstrated by Google, Honeywell and others using well-known algorithms and encodings.
Important concepts and algorithms impacting QEC have emerged from the HEP theory community, including aspects of the AdS/CFT dual correspondence.

An important and yet unanswered question is whether meaningful predictions can be extracted from quantum simulations performed on NISQ-era devices.  While it is expected that qualitative or conceptual advances can be made in certain areas, such as fragmentation,
during the NISQ-era, precise quantitative predictions for comparison with experiment will almost certainly be the perview of QEC-devices.  Our community is working toward answering this crucial question.

In quantum simulations and sensing for the HEP science mission, QEC will play a critical role in advancing the science by greatly extending the capabilities of computing and sensing devices, and is key to providing  precise and reliable results.
In developing HEP applications, the overheads that will be required for QEC should be included in estimating required resources and expected uncertainties. These will be different for different architectures, for instance between trapped-ion systems, Rydberg systems and SRF-cavity systems.
This is a crucial line of work that requires robust support for the theory community in HEP in order to undertake it in a timely fashion --- it is important to advance the effort now.

Unlike for exascale computing requirements, our field is only at the earliest stages of assessing quantum resource requirements for accomplishing any given objective. 
The next 5 years or so is the period where such estimates acquire improved reliability.
It is already the case that error mitigation is being utilized to find results from NISQ devices, including in the measurements, in extrapolating CNOT errors,  and ensemble methods to reduce deviations from gauge-invariant space.  Devices are now available where measurements can be interspersed during evolution, as part of error mitigation, which is enabling some protocols to be evaluated.

%%%%%%%%%%%%%%%%%%%%%%%
\section{Science Need for Advanced Calculations of Quantum Materials in HEP}
\label{sec:quantummaterials}
\noindent
{\it Note : The contents of this section are, in part, taken and modified from the Snowmass Whitepaper: {\bf Computational needs of quantum mechanical calculations of materials for high-energy physics}~\cite{Griffin:2022qlx} }
\vskip 0.1in

Next-generation HEP experiments are firmly entering the `quantum' realm where the energy scales and interactions of interest are approaching quantum limits.
%~\cite{Farhi_et_al:2015, Fradkin_et_al:2015, Chattopadhyay_et_al:2016}.
This is aptly exemplified in searches for new physics (such as dark matter (DM)) that lie below the weak mass scale both in proposing new modes of coupling to the standard model, and in strategies for sensors and readout that surpass current limits. 
Therefore, an in-depth understanding of the quantum behavior of the \textit{materials} that make up target/sensing systems is now needed for:
\begin{itemize}
  \item Accurate calculations of the material's form factor for specific couplings to new physics,
  \item Target/sensing efficiency calculations that incorporate decay channels and carrier transport,
  \item The role of materials inhomogeneities, defects and other decoherence channels for interpretation of near-threshold measurements,
  \item Materials design for the systematic improvement of detector/sensor regimes,
  \item Novel detection/sensing platforms based on (exotic) materials properties.
\end{itemize}
This is a resource-intensive computing problem that must be properly resourced to drive progress in quantum science in HEP.

First-principles descriptions of such materials' properties are possible using state-of-the-art Density Functional Theory (DFT) calculations.
%~\cite{Hafner_et_al:2006}.
DFT is the work-horse for the \textit{ab initio} description of materials at a quantum mechanical level and has recently been applied to accurately describe a range of HEP science problems such as DM-electron and DM-phonon interactions in a variety of materials.
Being entirely first-principles these robust methods eliminate the need for empirical parameters, and can describe and predict the behavior of materials.
The power of these approaches can also enable the inverse design of target functional properties.

Applications of \textit{ab initio} approaches in HEP include the prediction of novel materials systems/phenomena for targets and sensors including new chemistries, combinations of materials, and material architectures that can both maximize current detector/sensing efficiency and propose entirely new systems with improved functionality.
They can also quantify material-based origins of excess signals, diagnose decoherence in quantum systems, and suggest mitigation strategies --- e.g., these methods recently identified the origin of spin-based decoherence channels in superconducting qubits, and the improvements led to a factor of 6 increase in resonator quality factor%\cite{Altoe_et_al:2022}.
Current applications of quantum materials in HEP focus on single-particle phenomena that are typically described in a non-interacting framework.
However, these proposals do not take advantage of the thriving field of correlated quantum phenomena where small perturbations can result in threshold or cascade events, apt for small mass/energy detection.%\cite{Keimer/Moore:2017}.
Looking to the future, contemporary discoveries in quantum materials of novel correlated phases and phenomena including spin, topology and orbital degrees of freedom should be explored for HEP applications, with the corresponding quantum mechanical calculations going beyond DFT to accurately include strongly-correlated physics.

%%%%%%%%%%%
\subsection{Computational Efforts in Quantum Materials for HEP}

Calculations of quantum materials are currently used to estimate electronic and phononic properties of materials for HEP applications.
%~\cite{Griffin:2019}.
First-principles methods for these calculations offer a balance of accuracy and efficiency, and are extensible to higher-order diagrammatic expansions of quasiparticle interactions (e.g. multiphonon responses). 
However, computationally expensive, beyond-DFT approaches are required when phenomena beyond their mean-field, ground state properties are relevant. 
The accurate description of quasiparticle excitations requires calculation of the self-energy of the many-body electron problem through the GW Approximation, resulting in increased computational cost.
Strong correlations beyond those captured with simple extensions to DFT can be incorporated through the construction of embedding theories such as Dynamical Mean-Field Theory (DMFT), which are again expensive.
Future applications of quantum computers include simulations of such quantum embedding models (e.g. the Hubbard model). 

Materials theory and design is enhanced through materials databases (the Materials Genome Initiative) facilitated by both improved quantum mechanical methodology and HPC capabilities.
Using materials informatics approaches, novel chemistries and materials can be identified from these large databases (e.g. the Materials Project\cite{MP}) by combining high-throughput calculations, materials theory, data science, and machine learning, and have made key discoveries in fields as disparate as multiferroics, superconductivity and photovoltaics.
Nascent efforts to use such computational materials informatics approaches are being explored for novel detector materials.
Material predictions can then be used as a guide for the synthesis of optimal sensor and detector materials, with a theory/experiment feedback loop for improved predictions and synthesis.
Automation of high-throughput calculations and searches can accelerate discovery through workflows developed for materials theory applications, towards optimization of target and response properties. 
These results can be made available to the community through online databases for general reference datasets.

\subsection{General Patterns of HPC Usage}

The most used quantum mechanical first principles method is DFT which, for example, accounts for over 70\% of NERSC allocation time in materials science/solid state physics, and is the most used algorithm at NERSC.
A standard DFT calculation solves the many-body Schr\"{o}dinger equation using careful approximations -- most common DFT algorithms have O(N$^3$) scaling where N is the number of electrons in the system.
Therefore, computational requirements can reach very large scales with increasing system sizes and/or complexity (e.g. accurate simulations of interfaces of multi-component systems).
While a typical calculation solves the many-body Hamiltonian for total energy in a single self-consistent calculation, more complex calculations needing several self-consistent calculations and/or derivatives of energy to obtain forces (e.g. to find optimal atomic geometries) can rapidly increase the computational resources needed.

Workflows for calculations of materials make use of a wide variety of HPC resources, ranging from few-node local clusters and mid-sized institution compute clusters to larger scale national facilities such as NERSC and XSEDE. In addition to conventional CPU architectures, several standard code packages are now optimized and scale well with GPUs.
Most practitioners do not use any of these resources exclusively with typical workflows requiring a combination for file preparation and testing (local), and larger production runs (national HPC facilities). While some analysis can be performed on-the-fly, most workflows will make extensive use of a combination of HPC resources for preparation, analysis and curation of datasets. Because of this, efficient transfer of large data sets between these different compute resources is needed both during production runs and following the completion of the project for storage and archiving (often datasets may not reside on the HPC facilities where they have been created for future access, curation and re-use). 

Workflows using DFT and beyond-DFT methods to calculate materials properties typically generated 10s-100s TB of data per user, depending on the complexity of the system and the targeted properties.
Much of this will be high-quality materials data that can be used for a wide range of applications both within and outside of HEP.
To avoid duplication of efforts, and to maximize the utility of these computationally demanding calculations, they should be disseminated and maintained in databases that can be easily accessed and interpreted across disciplines, and it is important to support efforts to accomplish this task, e.g. \cite{MDF,MPContribs,MLExchange}.

%%%%%%%%%%%%%%%%%%%%%%%
\section{Quantum networks research support}
\label{sec:quantumnetworks}
Quantum networks represent interacting collections of quantum computers and quantum sensors that amplify the benefits of any individual elements. This arises from the entanglement that may be generated across the network as well as the integration of different functional elements. Currently, there is a growing effort to demonstrate quantum network functionality supported by research and development of individual quantum devices and methods for transmitting and transducing information between them.
We recommend a Snowmass Whitepaper \cite{https://doi.org/10.48550/arxiv.2203.16979} and the DOE Quantum Internet Blueprint \cite{osti_1638794} for further details.
\par 
A leading example of the benefit afforded by quantum networks for HEP is the case of precision timekeeping. Atomic clocks are perhaps one of the best develop quantum technologies with a precision to within a fraction of second over the lifetime of the Universe. Recent work has shown that a quantum network of atomic clocks can result in a substantial boost to the overall precision if multiple clocks are phase locked and connected by entanglement. The world’s most accurate atomic clock has reached a frequency accuracy of $10^{-21}$, and higher precision has the potential to transform global timekeeping, enabling orders-of-magnitude improvements in measurement accuracy and sensor resolution. 
\par 
For example, geographically-distributed networks are paramount in searches for dark matter transients and exotic field bursts from  powerful astrophysical events such as black hole mergers. Spatially extended dark matter may couple to the Standard Model and induce transient changes in fundamental constants,
such as the fine structure constant. The non-local correlations generated within a quantum network of atomic clocks may be used to cross-correlate signals from different times and locations. As of now, there are no searches with a cross-node entangled network of atomic clocks, as such networks are in their infancy.
\par 
Quantum networks of magnetometers offer similar advantages to boost sensitivity and enable exploration of new signals. Ultra-sensitive detection of magnetic fields has many applications, including the detection of WIMPs. For these networks, individual atomic magnetometers have reached parity with SQUIDS (superconducting quantum interference devices) to demonstrate sensitivities below 1 fT/pHz. 
\par
Within the context of astrometry, optical interferometry has played an important role for detection and characterization. Whereas traditional Michelson interferometers have been developed at the single-photon level, the extension to multi-photon entangled states offers new opportunities to enhance sensitivity to the microarcsecond. The resulting quantum-assisted telescopes would be capable of using non-local correlations in the temporal and spatial characteristics of light.  
\par
There is a connection between quantum networks and quantum machine learning (QML, discussed in Section \ref{sec:quantumdataana}).
In particular, while QML and other quantum analysis algorithms often have challenges operating on classical data due to input-output constraints, even in the NISQ-era it may prove advantageous to utilize QML on the intrinsically entangled states utilized in quantum networks.
This is an area worthy of additional study.
\par
The quantum networking applications afforded by entangling a variety of quantum devices together faces several key technical challenges. Methods for connecting, interacting, storing and relaying quantum information through the network are needed, and this often requires transducing information across quantum media. Transduction converts the encoding of quantum information from one physical form into another form. Transduction requires sources of quantum states, where fidelity and repetition rate are key metrics for assessing performance. The corresponding methods for detecting such states are required too. In many quantum networks, multi-photon entangled states serve a key role for mediating interactions as the information encoded in light can be transported quickly. Transduction is then an essential step in distributing these quantum states. In addition, quantum memory represents a class of methods to store intermediate states of quantum information that enable quantum repeaters. The latter is a means by which the network may correct errors during distribution of quantum states.
Transduction, quantum memory, and quantum repeaters are essential research priorities for expanding the transmission distance and scaling the performance of quantum networks. 

%%%%%%%%%%%%%%%%%%%%%%%
\section{Quantum computing for data analysis research support}
\label{sec:quantumdataana}

Quantum computing offers a number of potentially interesting approaches to data analysis in HEP.
While the data-input output problem looms over all the applications of quantum computers for analysis of classical data, the opportunities to explore a ``post-Moore's Law'' must be properly investigated.
Quantum simulation is, of course, able to contribute to data analysis by enabling better tools for creating synthetic data for designing experiments and analyses, but these applications are largely covered in the discussions on quantum simulation previously in this whitepaper, at least with respect to traditional Monte Carlo (MC) inputs like matrix elements, etc.
But beyond being a tool for better MC, quantum computing may offer a number of benefits to the direct analysis of HEP data.
These issues are discussed in Snowmass Whitepapers \cite{humble2022advancinghep,Delgado:2022tpc}.

Many of the most famous examples of quantum advantage are in solving linear algebra problems, and these may accelerate data analysis tasks when we are fully in the error-correction era of quantum computing.
In the interim, NISQ-era applications focus more on machine learning (ML).
Quantum machine learning (QML) faces many challenges, particularly because ``classical'' ML benefits from enormous industrial investments and research, making the quest for quantum advantage particularly difficult --- see also \cite{schuld2022quantadvgoal} for a thoughtful discussion of whether quantum advantage is the proper goal of QML research today.
A major problem in evaluating QML is the frequent need to compress the data in order to represent events on modern quantum hardware.
This necessitates using the same compression for classical benchmarks (which prevents us in many cases from using the best classical benchmarks).
It also makes it challenging to understand qubit scaling because techniques like Principle Component Analysis often provide marginally less and less information for each additional value allowed in the compressed state.
Despite these challenge, quantum ML has offered applications that are competitive with strong classical techniques and has offered new techniques that may offer enabling insights, both into HEP data problems and (Q)ML theory.

The Whitepaper  in Ref.~\cite{Delgado:2022tpc} 
provides a thorough review of QML-based applications in HEP.
There are already a large number of published studies in interesting potential application areas, including Track reconstruction, Jet reconstruction for $e+/e-$ events, Event classification, Data generation and augmentation, and Quantum-inspired algorithms.
Generally, researchers are able to find approximate parity between quantum algorithms and at least one classical benchmark.
In a few cases, researchers claim to have found quantum algorithms that outperform the classical benchmark, but it is hard to evaluate these claims because (1) it is not possible to understand how rigorous the optimization of the classical benchmark was and (2) all of the algorithms were conducted on a system of qubits modest enough to replicate with exact simulation, meaning they were effectively classical machine learning algorithms that utilized a simulation of some level of entanglement and superposition in its calculations.
The main missing piece in evaluating this second class of algorithms is understanding how performance scales with the number of qubits.
If the classifiers are able to demonstrate scaling improvements with more qubits, we will have candidate applications for beyond-classical performance.
More work and research is required in this promising area.

The areas with most immediate near-term potential for real scientific impact may be in generative models and quantum-inspired algorithms.
Even on NISQ machines, quantum computers have demonstrated the ability to sample from probability distributions that prohibitively expensive on classical computers.
While we typically lack theoretical guarantees, it is intuitively compelling that the ability to encode more sophisticated probability distributions in large Hilbert spaces should lead to performance advantages in generative modeling.

Perhaps the area where HEP has the most to offer to QML research that is unique is in the utilization of quantum sensor data, e.g., from dark matter searches \cite{https://doi.org/10.48550/arxiv.2203.05375,https://doi.org/10.48550/arxiv.2203.12714}.
The technological problem of linking sensors to a quantum computer is highly non-trivial, but conceptually, this is an application area where HEP may be uniquely suited to make contributions of interest on the scale of the national quantum ecosystem.

\emph{Quantum inspired algorithms} are also discussed in the \cite{Delgado:2022tpc} Whitepaper.
``Dequantization'' \cite{etangthequantumdestroyer,dequantlearning} has shed light on a variety of improvements to classical algorithms of broad impact in HEP and is an area worth investigating further.
In a concrete sense, these are contributions quantum computing has already made to our computational capabilities. 

%%%%%%%%%%%%%%%%%%%%%%%
\section{Quantum tools and testbeds}
\label{sec:quantumtoolsandtestbeds}
Alongside access to quantum computing systems, there is a need to develop software tools that address the use of quantum computers for HEP specific problem sets \cite{humble2022advancinghep}. Such tools are needed, in part, to facilitate access and manage the complexity of using quantum devices. Methods for expressing quantum algorithms clearly and concisely as well as reducing these representations to efficient forms that yield accurate results are essential. All efforts to apply quantum computing in HEP to date have consisted of adapting established general-purpose quantum algorithms and tools to HEP-specific tasks. Most demonstrations have involved only toy problems, and scaling up these ideas to large problem sets will require more sophisticated tools. In addition, the legacy of HEP tools available will require integration of quantum computing with existing workflows.
\par
In addition to the software tools needed for programming quantum computers, there is a clear need for tools that evaluate how specialized processing may be applied to HEP problem sets. Quantum computing supports both digital and analog variations, and there are many different technologies under consideration to realize these models. The presence of noise and errors in these technologies make fault-tolerant designs of the hardware and software an important consideration in building these specialized devices. The collaborative design, or co-design, of applications that run on these systems will be essential to success as appreciation for the error correction codes, hardware architectures, compiler optimizations, and problem structure are all required to build an efficient solution. This is especially important for minimizing the time to solution when operating fault tolerantly.
\par 
While considerations for codesign may be applied to many applications of quantum computing, each requires tailored approaches. For example, mapping HEP problems into the fault-tolerant era comprises a large and important set of tasks. Specific encodings that preserve gauge invariance may allow for more highly optimized representations and understanding how to develop these requirements alongside error-correction schemes with efficient overheads is unlikely to be addressed outside of the HEP community. These hardware-software co-design efforts must engage domain scientists with specialists from computer science and math as a community-led strategy. Cutting-edge HEP applications will drive fundamental progress in computer science and math that have spill-over effects into other fields. Similarly, by remaining engaged in cross-disciplinary projects, the HEP community will benefit more quickly and efficiently from advances originating in other areas.
\par
Software tools, including programming languages, compilers, debugger, and profilers are needed broadly within the burgeoning quantum computingg community, and many tools may be addressed by a generalized research agenda. However, there are critical needs for the HEP community reflecting the data types and encodings that are likely to influence tool development. Therefore, it will be important to maintain close coordination. Similarly, deep involvemnet in on-going hardware testbeds will provide input on the capabilities and features offered by quantum computer themselves, which will help guide application development. Robust participation in testbed programs will help HEP answer the questions around access and whether we need specialized, dedicated hardware, and how to best execute low-level control.
\par
Key feedback on the utility of these tools is there ability to advance toward quantum computational advantage, as measured relative to conventional methods for solving the same problems. Monitoring progress toward quantum advantage will require methods for tracking improvements in key metrics, including time-to-solution and application accuracy. Benchmarking quantum devices using HEP problem sets has seen modest developments but this is far from wide-spread and standardized adoption. In addition, quantum-inspired algorithms have shown promise for improving conventional methods.

%%%%%%%%%%%%%%%%%%%%%%%
\section{Conclusions}
\label{sec:conclusions}

Quantum computing is poised to make major advances to the core HEP science mission.
We know this to be true because (1) we have identified computational problems that are classically intractable, but that would revolutionize HEP if we could solve them, and (2) we already know that Quantum Computers provide exponential speedup over classical computers for some of these problems.
We are not short on applications of impact in HEP. 
There may be more interesting problems --- in fact, there almost certainly are! --- but we have enough now to compel action.
We have a set of clear targets that will only be achieved with substantial research and development.

This is a formative moment in the development of both HEP and QIS  research, especially when considered together.
The transition from the current NISQ-era of QIS 
to an era of fault-tolerant, quantum error corrected computation
will revolutionize the utility of quantum computers. Over the next decade, these technical advances in QIS will have profound impacts on technology, science, the economy, and society itself.
HEP has a useful role to play in advancing quantum computing to this new age and also stands to benefit enormously from an active engagement in QIS.
It is important to engage during the NISQ era in order to accelerate the path to fault tolerance and to ensure we will be able to fully utilize error corrected quantum computers when they become available.
A deep and active research engagement in quantum computing physics, devices, and algorithms will enable progress on HEP science problems that may never be solved without quantum computing.
This expansion of the HEP research portfolio will make meaningful contributions to one of the cornerstone technologies of the Twenty-First Century.
We should not view QIS as an excursion from traditional HEP science, but instead we should see it for it is --- a tool for fundamental physics that is fully part of our science mission.

Delivering these results requires a HEP program that is broadly engaged with many different quantum technologies and research areas, and simultaneously deep in those areas where HEP offers the most to the national quantum ecosystem --- in devices, superconducting technology, controls, theory, algorithms, and applications.
It is absolutely crucial the HEP community build the research infrastructure to become a developer of QIS technologies, and not a simple consumer, in order to benefit from quantum science and to ensure access to quantum devices capable of meeting the goals of a broader science program.
Given the scale of the global quantum ecosystem and investments  made in quantum computing outside of HEP, a strong partnership with industry and other domain sciences will ensure an agile position to capitalize on breakthroughs.
As the broader QIS communities continues to develop scalable fault-tolerant approaches to error-corrected quantum computing, these partnership will mitigate technical risk and offer programmatic flexibility.
At the same time, we cannot rely on the efforts of other domain science fields in QIS to solve HEP problems any more than we may rely on them to solve our problems with classical computing.
It is true quantum chemists will do a great deal to advance QIS, but the same is true for high performance classical computing and we have never been in a position where we simply wait for them to solve our research problems.
We may benefit from their experiences, but if we want to solve the most challenging computational problems in HEP, we must do it ourselves.

The grand challenges that may be addressed with error-corrected quantum computing center on the simulation of quantum systems,
which almost certainly to push lattice gauge theory toward its full potential.
Furthermore, there are profound questions about the nature of reality and the connection of information theory with fundamental physics that may be impossible to answer without a quantum computer.
Is a quantum computer needed to simulate our universe?
There are strong reasons to believe so --- but it would be equally fascinating and surprising if this were not true!
Perhaps a quantum computer itself is insufficient to truly simulate physics and a quantum-gravitic computer is required?
Studying these questions should be part of the core HEP mission for the next decade.

This is the first Snowmass planning activity to consider QIS-HEP research outside of formal theory activities. The development of a QIS-specific research plan is on a path to become a major component of the HEP research portfolio during the next decade, with mutually beneficial advances to both HEP and the nation's broad and rapidly growing QIS program.  
The integration of QIS with existing and other new HEP research activities should be deliberate, with continuing assessments of expected resource requirements.
A robust QIS base program within the HEP research portfolio should continue to support new initiatives through small grant programs and focused project portfolios like the DOE QuantISED program, while
 enthusiastically supporting the large-scale research projects enabled by national QIS research centers.

%%%%%%%%%%%%%%%%%%%%%%%
\section{Acknowledgements}

We are extremely grateful to all who contributed to writing the Whitepapers from which this document is derived.
We also thank the multitude of colleagues who reviewed drafts of this document and provided valuable feedback, including: Christian Bauer, Ignacio Cirac, Andrea Delgado, Sinéad Griffin, Yannick Meurice, Benjamin Nachman, James Osborn, Nicholas Peters, Panagiotis Spentzouris, Sofia Vallecorsa, and Silvia Zorzetti.

%% Gabe stuff...
This document was prepared using the resources of the Fermi National Accelerator Laboratory (Fermilab), a U.S. Department of Energy (DOE), Office of Science, HEP User Facility. 
Fermilab is managed by Fermi Research Alliance, LLC (FRA), acting under Contract No. DE-AC02-07CH11359.
%% Travis ACK statement
This material is based upon work supported by the U.S. Department of Energy, Office of Science, National Quantum Information Science Research Centers, Quantum Science Center. This manuscript has been authored by UT-Battelle, LLC under Contract No.~DE-AC05-00OR22725 with the U.S.~Department of Energy. 
%% Savage
This work was supported in part by the U.S. Department of Energy,
Office of Science, Office of Nuclear Physics, InQubator for Quantum Simulation (IQuS) under Award Number DOE (NP) Award DE-SC0020970 (Savage).

\newpage

\bibliographystyle{apsrev4-2}
\bibliography{references}

\end{document}

%% file: new_executive_summary.tex
%
% Aim for 1 page only!
% 
Quantum Information Science (QIS) has the potential for transformative impacts on High Energy Physics (HEP).
Quantum computing specifically is part of a trinity of technologies --- sensing, networks, and computing.
Although the science mission of HEP stands to benefit from strongly from all three, particularly when they are working in tandem, this report will primarily focus on quantum computing as a piece of the ``Computational Frontier'' in HEP.

% --- What is the physics grand challenge we want P5 to focus on?
% Need to make this stronger... what is the quantum/entanglement frontier pillar we want highlighted in strong language?
Quantum computers are widely misunderstood.
They are not general-purpose devices for speeding up calculations; rather they are tools for discovery, much like particle accelerators or telescopes.
The HEP science program is engaged with a number of crucial challenges that will almost certainly require fault-tolerant quantum computers to solve.
Solving quantum many-body simulation will profoundly impact our practical capabilities in quantum field theory, and revolutionize our understanding of the world around us.
Quantum computers are also expected to have dramatic impacts on the capabilities of event generators, and they are also poised to serve as novel data analysis machines, especially when coupled to quantum sensing experiments.
We also hope to use quantum computers for some of the most fundamental questions one may ask about nature --- is a quantum computer necessary to efficiently simulate the physics our universe? 
Or is something more required?

% --- What is the concise set of recommendations for focus areas?
% \item Hardware exploration
% \item Co-design leveraging HEP expertise
The hardware roadmap from present day machines to fault-tolerant quantum computers is not clearly marked --- multiple different technologies remain strong contenders.
Therefore we must strike a balance between testing a wide variety of platforms by building partnerships that leverage HEP expertise and co-design efforts aimed at building unique computational platforms expressly designed to solve HEP science problems.
In both pieces of this overall strategy, HEP will advance its science mission while also making impactful contributions to the national QIS ecosystem.
It is important to leverage all that HEP has to offer to QIS both to enable the next generation of fundamental physics research, and to secure our role in this rapidly growing research endeavor.
% \item Theory investments in quantum simulation
% \item Workforce
We need to develop and leverage quantum error correction (QEC) codes for physics problems and invest in better understanding the connections between field theory and quantum computing.
Finally, we must look to the future and both develop QIS expertise within HEP institutions and play a strong role in training an inclusive quantum computing workforce.

\clearpage

%% file: recommendations.tex
The most important facts for framing quantum computing research in HEP are:
\begin{itemize}
    \item There are computational problems that are completely intractable classically, but which would revolutionize HEP if they could be solved (simulating the dynamics of a QFT is an example).  
    \item We already know that Quantum Computers provide exponential speedup over classical computers for some of these calculations.
\end{itemize}
We are not searching for applications of impact in HEP. Instead, we have very clear targets that require substantial research and development.

In light of this, we recommend:
\begin{enumerate}
    \item The hardware roadmap to fault-tolerant machines that are well-suited for HEP problems is not clearly marked. Therefore:
    \begin{itemize}
        \item We must invest in a diversity of hardware platforms available across the Quantum Information Science (QIS) landscape.
        \item We must invest in research for building quantum hardware matched to our requirements. HEP problems are different than those the commercial platforms are optimizing for. We have a strong history (e.g. lattice QCD) for effective co-design in a computational domain where physics-informed development may be brought to bear.
    \end{itemize}
    \item QIS is a large field that couples to many domain science disciplines and the economy as a whole. Therefore:
    \begin{itemize}
        \item We must leverage HEP strengths in superconducting technology, physics domain knowledge, fast electronics, cryogenic electronics, system architecture, integration, and engineering, and large scale scientific project management to advance QIS at the broadest level in order to fully justify HEP participation in this rapidly growing area.
        We should invest robustly in these areas to support further advancements.
        \item We must engage with the broader QIS community will enable us to leverage the large investments being made elsewhere to enable faster and more robust progress on the HEP science program.
    \end{itemize}
    \item A ``trinity'' of technologies comprise modern quantum science: computing, sensing, and networks. Therefore:
    \begin{itemize}
        \item We must invest in all three of these technologies and in the synergies between them. All three are important for each other and all three enable important pieces of the HEP science program.
        \item We must leverage HEP expertise in the technologies underpinning quantum networks, both for the quantum internet more broadly but especially for the potential impacts in computing and sensing. Distributed sensors entangled through a quantum network could be of great interest to HEP precision experiments.
        Quantum computing and sensing technologies often develop together and in this sense quantum computing efforts may directly impact novel new physics searches.
    \end{itemize}
    \item Workforce development is another key area of investment. Therefore:
    \begin{itemize}
        \item We must support the significant theoretical work that must be done to fully address the potential of quantum simulation for HEP on fault tolerant computing.
        \item We must develop long-term jobs in the field for people working at the intersection of QIS and HEP.
        \item We must play a role in fostering a diverse and inclusive workforce that gives all segments of society a say in how this research shapes our lives.
    \end{itemize}
\end{enumerate}

\clearpage